\documentclass[12pt]{article}

\usepackage{graphicx}
\usepackage{subcaption}
\usepackage{amsmath}
\usepackage{dsfont}
\usepackage{amssymb}
\usepackage{abstract}
\usepackage{hyperref}
\usepackage{parskip}
\usepackage{enumerate}
\usepackage{fancyhdr}
\usepackage[xcdraw,table]{xcolor}
\usepackage[round]{natbib}
\usepackage{multirow}
\usepackage{algorithm}
\usepackage{algorithmicx}
\usepackage{algpseudocode}

\usepackage[margin=0.95in]{geometry}
\usepackage{fancyvrb}

\newcounter{probnum}
\allowdisplaybreaks

\definecolor{tabblue}{rgb}{.870588,.905882,.94902}
\definecolor{gray}{rgb}{0.5,0.5,0.5}
\definecolor{black}{rgb}{0,0,0}
\definecolor{white}{rgb}{1,1,1}
\definecolor{blue}{rgb}{0.0,0.0,1}
\newcommand{\inblue}[1]{\color{blue}#1\color{black}}
\definecolor{green}{rgb}{0,0.5,0}
\newcommand{\ingreen}[1]{\color{green}#1\color{black}}
\definecolor{yellow}{rgb}{1,0.549,0}

\definecolor{red}{rgb}{0.6,0.0,0.0}
\newcommand{\inred}[1]{\color{red}#1\color{black}}
\definecolor{darkred}{rgb}{0.9,0.4,0}
\newcommand{\indarkred}[1]{\color{darkred}#1\color{black}}
\definecolor{purple}{rgb}{0.58,0,0.827}

\definecolor{backgcode}{rgb}{0.97,0.97,0.8}
\definecolor{Brown}{cmyk}{0,0.81,1,0.60}
\definecolor{OliveGreen}{cmyk}{0.64,0,0.95,0.40}
\definecolor{CadetBlue}{cmyk}{0.62,0.57,0.23,0}

\DeclareMathOperator*{\argmin}{arg\,min~}

\newcommand{\bv}[1]{\boldsymbol{#1}}

\newcommand{\sigsq}{\sigma^2}
\newcommand{\sigsqe}{\sigma^2_e}

\newcommand{\tausq}{\tau^2}

\newcommand{\ybar}{\bar{y}}
\newcommand{\Ybar}{\bar{Y}}
\newcommand{\YbarT}{\Ybar_T}
\newcommand{\YbarC}{\Ybar_C}
\newcommand{\xbar}{\bar{x}}

\newcommand{\iid}{~{\buildrel iid \over \sim}~}

\newcommand{\half}{\frac{1}{2}}

\newcommand{\X}{\bv{X}}
\renewcommand{\S}{\bv{S}}
\newcommand{\Sinv}{\S^{-1}}

\newcommand{\x}{\bv{x}}

\newcommand{\reals}{\mathbb{R}}

\newcommand{\limitn}{\lim_{n \rightarrow \infty}}

\newcommand{\beqn}{\vspace{-0.25cm}\begin{eqnarray*}}
\newcommand{\eeqn}{\end{eqnarray*}}
\newcommand{\bneqn}{\vspace{-0.25cm}\begin{eqnarray}}
\newcommand{\eneqn}{\end{eqnarray}}
\newcommand{\benum}{\begin{enumerate}}
\newcommand{\eenum}{\end{enumerate}}

\newcommand{\parens}[1]{\left(#1\right)}
\newcommand{\squared}[1]{\parens{#1}^2}
\newcommand{\angbrace}[1]{\left<#1\right>}

\newcommand{\bracks}[1]{\left[#1\right]}
\newcommand{\braces}[1]{\left\{#1\right\}}

\newcommand{\expe}[1]{\mathbb{E}\bracks{#1}}
\newcommand{\cexpe}[2]{\expe{#1\,|\,#2}}

\newcommand{\indic}[1]{\mathds{1}_{#1}}
\newcommand{\var}[1]{\mathbb{V}\text{ar}\bracks{#1}}

\newcommand{\se}[1]{\text{SE}\bracks{#1}}

\newcommand{\oneover}[1]{\frac{1}{#1}}
\newcommand{\overtwo}[1]{\frac{#1}{2}}

\newcommand{\bernoulli}[1]{\mathrm{Bern}\parens{#1}}

\newcommand{\normnot}[2]{\mathcal{N}\parens{#1,\,#2}}

\newcommand{\stdnormnot}{\normnot{0}{1}}

\newcommand{\SigField}{\mathcal{F}}

\newcommand{\convp}{~{\buildrel p \over \rightarrow}~}

\newcommand{\convd}{~{\buildrel \mathcal{D} \over \longrightarrow}~}

\newcommand{\pval}{p_{\text{val}}}

\newcommand{\betaT}{\beta_T}
\newcommand{\epsilonrv}{\mathcal{E}}
\newcommand{\indicTi}{\indic{T,i}}

\newcommand{\nr}{n_{R}}
\newcommand{\nrc}{n_{R,C}}
\newcommand{\nrt}{n_{R,T}}

\newcommand{\Dbar}{\bar{D}}
\newcommand{\SsqDbar}{S^2_{\Dbar}}
\newcommand{\SsqR}{S^2_R}
\newcommand{\YbarRT}{\Ybar_{R,T}}
\newcommand{\YbarRC}{\Ybar_{R,C}}
\newcommand{\YbarRTMinusYbarRC}{\YbarRT - \YbarRC}

\newcommand{\dbar}{\bar{d}}
\newcommand{\ssqDbar}{s^2_{\Dbar}}
\newcommand{\ssqR}{s^2_R}
\newcommand{\ybarRT}{\ybar_{R,T}}
\newcommand{\ybarRC}{\ybar_{R,C}}
\newcommand{\ybarRTMinusybarRC}{\ybarRT - \ybarRC}

\newcommand{\xnew}{\x_{\text{new}}}
\newcommand{\xold}{\x_{\text{old}}}
\newcommand{\Xnew}{\X_{\text{new}}}
\newcommand{\Xold}{\X_{\text{old}}}

\newcommand{\sigsqR}{\sigsq_R}
\newcommand{\sigsqDbar}{\sigsq_{\bar{D}}}
\newcommand{\sigsqDeltaYbar}{\sigsq_{\Delta\Ybar}}

\title{Matching on-the-fly in Sequential Experiments for Higher Power and Efficiency}
\author{Adam Kapelner and Abba Krieger}

\begin{document}
\maketitle

\begin{abstract}
We propose a dynamic allocation procedure that increases power and efficiency when measuring an average treatment effect in sequential randomized trials. Subjects arrive iteratively and are either randomized \textit{or} paired via a matching criterion to a previously randomized subject and administered the alternate treatment. We develop estimators for the average treatment effect that combine information from both the matched pairs and unmatched subjects as well as an exact test. Simulations illustrate the method's higher efficiency and power over competing allocation procedures in both controlled scenarios and historical experimental data. 
\end{abstract}

\section{Introduction}

The gold standard of experimentation, randomization, is only golden with large sample size. With small sample size, the empirical distributions of relevant covariates can be different across treatments possibly masking an effect by creating bias and inflating variance. Some improvements over completely randomized design are rerandomization \citep{Morgan2012}, \textit{a priori} matching \citep{Raudenbush2007}, and \textit{adaptive design} which involves any change to the experiment or statistical procedures while the experiment is underway \citep{Chow2008}, even assigning subjects to different treatments upon failure or noncompliance \citep{Lei2012}.

We limit our focus to \textit{sequential experiments}, where subjects enter iteratively over time and the experimental condition is administered upon entrance. We develop a new adaptive design for sequential experiments whose goal is to elucidate an average treatment effect (ATE) between two treatments, which we call treatment (T) and control (C). Sequential experiments are very popular in both clinical trials and recently, crowdsourced-Internet experimentation \citep{Horton2011, Chandler2013}.

Our design is a new type of \textit{dynamic allocation}, a means of assigning T/C to newly-arrived subjects based on decisions about previous assignments, covariates, and/or responses \citep{HuRosenberger2006}. Proposals of dynamic allocation began with \citet{Efron1971}'s biased coin design. Here, a coin is biased in favor of the treatment with the fewest subjects, hence leading to better balance in treatment allocation. \citet{Wei1977}'s urn design generalizes this biased coin procedure, but both methods disregard the subjects' covariates.

The first line of defense to balance covariates is stratification (or ``blocking''), a classic strategy dating back to Fisher's agricultural experiments. Stratification becomes quickly impractical when the number of total covariate levels is large relative to sample size. \citet{Taves1974}, \citet{Wei1978b} and \citet{Begg1980} tackle these shortcomings by ``minimizing'' the imbalance between the treatments among all levels of covariates present. The most popular and widely implemented among these methods is \citet{Pocock1975} whose procedure involves picking a few arbitrary functions to tailor the imbalances. The selection of these functions is still an ongoing area of research \citep[for instance, see][]{Han2009}. Concerned by this arbitrariness, \citet{Atkinson1982} posits a method solidly rooted in linear model theory using $D_A$ optimality.


If the end goal of experiments is to find an effect, then the primary concern is estimator efficiency and test power. Stratification and minimization methods rely on the logic that greater balance among the covariates implies greater efficiency which is mathematically true \textit{only} in homoskedastic linear models \citep{Rosenberger2008}. $D_A$ optimality iteratively maximizes efficiency assuming the linear model without explicitly focusing on balance. Thus, we see one of the fundamental problems in previous allocation procedures is the reliance on the homoskedastic linear model, an assumption that is rarely true in practice. We wish to develop a dynamic allocation which is robust when the covariates combine non-linearly and with interactions to produce the response and no worse than current methods when the linear model holds.

The seminal guidebook, \citet{CookCampbell1979}, states ``whenever possible, it is recommended to minimize error terms'' and recommend \textit{matching} subjects before randomization on covariates to create better stratification. It was not a novel idea; \citet{Student1931} commented on the $n = 20,000$ children Lanarkshire Milk Experiment proposing the experiment should be performed exclusively on 50 identical twin pairs which would be randomly assigned T/C.

We propose matching iteratively, \textit{on-the-fly}. As subjects walk in the door or engage a survey online they should be matched with people ``similar'' to them who came in previously, a procedure which \citet{Whitehead1997} believes to be ``especially difficult.'' Imagine the following scenario of a trial testing whether a pill decreases blood pressure. The investigators determine that age, height, weight, and race should be collected as covariates as they are known to be related to blood pressure. Bob, 28, 5'10'', 180lb and white enters and is determined to fit the requirements of the study. By the flip of a coin, he receives the pill. The next day, Grace, 45, 5'2'', 105lb and Asian shows up. Based on inspection of this demographic data, she is clearly different from Bob; thus she is also randomized. Soon after, Joe, 29, 5'11", 185lb and white enters. We determine that he is similar to Bob, pair them, and deterministally administer to him the placebo. The trial continues and Mary would then await someone  to be matched with, which may or may not occur.

The algorithm is simple: incoming subjects are either randomized and placed in a holding pool, called the ``reservoir,'' or if they're found to match a subject already in the reservoir, they're matched and given their match's alternate treatment. The matches and the reservoir form two independent samples yielding two different estimators which are combined to estimate the ATE.

The closest idea we find in the literature is in \citet{Raghavarao1980} who computes the Mahalanobis distances between a new subject's covariates and the average covariates in the different treatment groups. The allocation to the treatment is then made via a biased coin with probabilities proportional to these distances. We use the idea of Mahalanobis distance which creates robustness to collinearity in the covariates, but we use it to match individual subjects together in pairs.

We layout our scenario assumptions, explain our algorithm and develop testing procedures in section \ref{sec:methods}. We demonstrate our procedure's improvements over previous procedures via simulations in section \ref{sec:simulations}. Our method performs particularly well in the case where the model is non-linear, performs respectably with linear models, and also performs respectively when the covariates do not inform the response. We then demonstrate higher efficiency using historical data from randomized controlled trials (RCT's) in section \ref{sec:real_data}, where the covariate-response model was unknown but most likely non-linear. We discuss and describe future directions in section \ref{sec:discussion}.

\section{The Algorithm, Estimation, and Testing}\label{sec:methods}

\subsection{Problem Formulation}\label{subsec:problem_formulation}

Subjects arrive sequentially and their covariates, denoted by $\x_i := \bracks{x_{i1}, \ldots, x_{ip}}$ which are either continuous or binary, are immediately recorded. The subjects must then be assigned to a treatment on-the-spot. We develop our method for allocating two treatments, T or C, denoted by the treatment indicator $\indicTi$. The response $y_i$ is continuous and can be collected at any time after allocation. We assume the following model with independent observations, an additive treatment effect, a possibly non-linear covariate effect, normal and homoskedastic errors, fixed covariate design, and sample size $n$ fixed in advance:

\bneqn\label{eq:response_model}
Y_i = \beta_T \indicTi + z_i + \epsilonrv_i, \quad\quad  z_i := f(\x_i), \quad \quad \epsilonrv_i \iid \normnot{0}{\sigsqe}, \quad \quad i \in \braces{1,\ldots,n}.
\eneqn

We wish to develop a dynamic allocation method followed by an unbiased estimator for $\beta_T$ with higher efficiency and thereby more powerful when testing a null hypothesis than previous approaches.

%

\subsection{The Algorithm}\label{subsec:algorithm}

The first few subjects enter the experiment and are randomized to T or C with the flip of a coin. These subjects comprise the ``reservoir.'' After a certain point, we would like to potentially match an incoming subject with subjects in the reservoir. We would like to match them on $f(\x)$, which is latent, so we match on what we consider is the next best thing, the $\x$'s themselves. We hope that $\x_1 \approx \x_2$ implies $f(\x_1) \approx f(\x_2)$ which is true if the function is sufficiently smooth. 

We match using squared Mahalanobis distance which gives a convenient scalar distance between points in $\reals^p$ adjusting for collinearities. This metric has a long implementation history in matching applications dating back to \citet{Rubin1979}. Matching using Mahalanobis distance and then randomizing the pairs to T/C has been demonstrated to result in better balance and higher power \citep{Greevy2004}. Further, the assumption of normal covariates seems to work well with real data even when the covariates are non-normal (see section \ref{sec:real_data}). After matches are produced, we do not make use of the normality assumption of the $\x$'s further in the delvelopment of the estimator. Improvements to the matching machinery that may be more robust to real-world covariate distributions are also discussed in section \ref{sec:discussion}. 

Thus, the new subject enters and the squared Mahalanobis distance between its covariate vector, $\xnew$, and each of the previous subject covariate vectors, the $\xold$'s, are calculated. Denote $\bv{S}$ as the covariates' sample variance-covariance matrix calculated with all subjects including the new subject. Assuming normal covariates, the squared Mahalanobis distance then has a scaled $F$ distribution given below:

\bneqn\label{eq:mahalanobis}
D_M^2 := (\Xnew - \Xold)^\top \Sinv (\Xnew - \Xold), \quad \quad \frac{n-p}{2p(n-1)} D_M^2  \sim F_{p,n-p}
\eneqn

We then take the minimum of the squared Mahalanobis distances between the new observation and each observation in the reservoir and calculate its probability. Let the minimum distance squared come from the previous subject, $\xold^*$. If the probability is less than $\lambda$, a pre-specified hyperparameter, then $\xnew$ and $\xold^*$ are matched together; if it's not, $\xnew$ is randomized and added to the reservoir. If $\xnew$ is matched, it is not added to the reservoir and $\xold$ is removed from the reservoir. $\indic{T,\text{new}}$ is then assigned to be $1 - \indic{T,\text{old}^*}$, i.e. the opposite treatment. The process is repeated until the $n^\text{th}$ entrant. We left out other implementation details in this discussion but make them explicit in algorithm \ref{alg:matching}. Note that our proposed procedure is considered a form of \textit{covariate-adaptive randomization} \citep[section 2]{Rosenberger2008} because we are using the covariates to determine the dynamic allocation.

\begin{algorithm}[htp]
\caption{The sequential matching algorithm for subjects entering the experiment. The algorithm requires $\lambda$ to be prespecified, which controls the ease of creating matches.}
\begin{algorithmic}[1]
\For{$t \gets \braces{1,\ldots,n}$} \Comment{$n$ is the total sample size, fixed a priori}
   \If{$t \leq p$ \textbf{or} reservoir empty}
	\State $\indic{T,t} \gets \bernoulli{\half}$ and $\bracks{\x_t, \indic{T,t}}$ is \textit{added} to the reservoir \Comment{randomize}
    \Else
	\State $\Sinv_t$ is calculated using $\x_1,\x_2, \ldots, \x_t$ \Comment{Estimate the true var-cov matrix}
	\State $T^{2^*} \gets\frac{p(t-1)}{t-p}F^*_{\lambda,p,t-p}$ \Comment{$F^*$ is the critical  cutoff from the $F$ distribution quantile}
	\ForAll{$\x_r$ in the reservoir}
		\State $T^2_r \gets \oneover{2} (\x_t - \x_r)^\top \Sinv_t (\x_t - \x_r)$
	\EndFor
	\State $T^2_{r^*} \gets \displaystyle\min_{r}\braces{T^2_r},~~r^* \gets \displaystyle \argmin_{r }\braces{T^2_r}$ \Comment{arbitrate ties if they exist}
	\If{$T^2_{r^*} \leq T^{2^*}$} 
		\State $\indic{T,t} \gets 1- \indic{T,r^*}$ \Comment{assign subject $t$ the opposite of $r^*$'s assignment}
		\State $\bracks{\x_{r^*}, \indic{T,r^*}}$ is \textit{removed} from the reservoir
		\State record $\angbrace{\x_t, \x_{r^*}}$ as a new match
	\Else
		\State $\indic{T,t} \gets \bernoulli{\half}$ and  $\bracks{\x_t, \indic{T,t}}$ is \textit{added} to the reservoir \Comment{randomize}
	\EndIf
    \EndIf
\EndFor
\end{algorithmic}
\label{alg:matching}
\end{algorithm}

Note that upon matching, the treatment indicator is assigned \textit{deterministically} to be the opposite of its match. This can cause selection bias if the investigator is not properly blinded.  In defense of our decision to make allocation deterministic for almost half the subjects, note that our algorithm is sufficiently complicated that a duplicitous investigator would not be able to guess whether the entering subject will be assigned T or C based on previous information during a clinical trial. We agree with \cite{Begg1980} that ``the idea that responsible investigators, even if they knew all the allocations to date, would spend their time playing games to try to guess a relatively complicated deterministic procedure... [is not] appealing.'' Additionally, \citet{McEntegart2003} discusses how this type of machination is unrealistic even in multi-center block permuted designs, which is much simpler than the allocation strategy proposed here. Further, if the procedure is implemented in an Internet-based experiment, the algorithm would be hard-coded and would not be subject to human tampering.

\subsection{Estimation and Hypothesis Testing}\label{subsec:estimator}

We first assume the covariate design matrix is fixed. If so, the subjects ultimately matched and the subjects ultimately found in the reservoir are fixed as well. Thus conditioning on the design is the equivalent to conditioning on the sigma field given below:

\beqn
\SigField = \sigma(\underbrace{\angbrace{\x_{T,1}, \x_{C,1}}, \angbrace{\x_{T,2}, \x_{C,2}}, \ldots, \angbrace{\x_{T,m}, \x_{C,m}}}_{\text{matched pairs}}, ~\underbrace{\x_{R,1}, \x_{R,2}, \ldots, \x_{R,n_R}}_{\text{reservoir}}).
\eeqn

Upon completion of the experiment, there are $m$ matched pairs and $n_R$ subjects in the reservoir, both quantities fixed since the sample size and the design are fixed ($n = 2m + n_R$). In our development of estimators and testing procedures, we always assume conditioning on $\SigField$, thus this notation is withheld going forward.

We focus on testing the classic hypotheses $H_0: \betaT = \beta_0$ versus $H_a: \betaT \neq \beta_0$. We consider testing under three model assumptions (a) the response has normal noise and may possibly depend on covariates but we do not wish to model their effect (b) the response has normal noise and depends linearly on covariates (c) the response has mean-centered noise and depends on covariates through an unknown model. We develop testing procedures for each situation: (a) a modification to the classic $\YbarT - \YbarC$ in section \ref{subsubsec:classic_test},  (b) a modification to ordinary least squares regression in section \ref{subsubsec:ols_test} and (c) an exact permutation test in section \ref{subsubsec:permutation_test}.

\subsubsection{The Classic Test}\label{subsubsec:classic_test}

We define $\Dbar$ as the estimator for the average of the differences of the $m$ matched pairs (treatment response minus control response) and $\YbarRT$, $\YbarRC$ as the estimators for the averages of the treatments and controls in the reservoir. We combine the estimators using a weight parameter, $B_T := w \Dbar + (1 - w) \parens{\YbarRTMinusYbarRC}$. We can find $w$ to minimize variance to obtain:

\bneqn\label{eq:estimator}
B_T = \frac{\sigsqR\Dbar + \sigsqDbar\parens{\YbarRTMinusYbarRC}}{\sigsqR + \sigsqDbar}, \quad\quad \var{B_T} = \frac{\sigsqR\sigsqDbar}{\sigsqR  + \sigsqDbar}.
\eneqn

$B_T$ is unbiased because $\Dbar$ and $\YbarRTMinusYbarRC$ are unbiased. Standardizing $B_T$ gives a standard normal due to the assumption of normal noise. To create a usable test statistic, note that the true variances are unknown, so we plugin $\SsqDbar$, the matched pairs sample variance estimator, and $\SsqR$, the pooled two-sample reservoir variance estimator:

\beqn
\SsqDbar &=& \oneover{m(m-1)}\sum_{i=1}^m \squared{D_i - \Dbar}, \\
\SsqR &=& \frac{\sum_{i=1}^{\nrt} \squared{Y_{R,T,i} - \Ybar_{R,T}} + \sum_{i=1}^{\nrc} \squared{Y_{R,C,i} - \Ybar_{R,C}}}{\nr -2} \parens{\oneover{\nrt} + \oneover{\nrc}}.
\eeqn

We use the notation $\nrt$ and $\nrc$ to be the number of treatments and controls in the reservoir ($\nr = \nrt+\nrc$). In practice, $\nrt$ is a random quantity, a binomial distribution with size $\nr$ and probability of one half. Howevere, we will assume $\nrt$ and $\nrc$ are fixed as an approximation. A more careful calculation could include the randomness of $\nrt$ and $\nrc$.

Equation \ref{eq:clt_true_var} shows the resulting statistic which has an asymptotically standard normal distribution since $\SsqDbar \convp \sigsqDbar$ and $\SsqR \convp \sigsqR$. Also, by the assumption of additive and normal noise, the estimator is also unbiased for finite $n$.

\bneqn\label{eq:clt_true_var}
\frac{B_T - \beta_0}{\se{B_T}} \approx \frac{\dfrac{\SsqR \Dbar + \SsqDbar \parens{\YbarRTMinusYbarRC}}{\SsqR + \SsqDbar} - \beta_0}{\sqrt{\dfrac{\SsqR\SsqDbar}{\SsqR + \SsqDbar}}}  ~\convd~ \stdnormnot
\eneqn

Note that in the case where there are no matched pairs, we default to the classic estimator and in the case where there are less than two treatments or controls in the reservoir, we default to the matched pairs estimator.

When is this estimator more efficient than the standard classic estimator, $\Delta\Ybar := \YbarT - \YbarC$? In other words, when is $\sigsqDeltaYbar / \sigsqDbar > 1$? Assuming perfect balance in its treatment allocation ($n_T = n_C = \overtwo{n}$) for the classic estimator and taking the expectation over both noise and treatment allocation, it can be shown that the variances are:

\bneqn\label{eq:var_sigsq_Dbar}
\sigsqDbar &=& \frac{1}{m^2}  \sum_{k=1}^m \squared{z_{T,k} - z_{C,k}} + \frac{2}{m} \sigsq_e, \quad\quad \sigsqDeltaYbar \approx \frac{4}{n^2} \sum_{i=1}^{n} z_{i}^2 + \frac{4}{n}\sigsqe.
\eneqn

This means that the better the matching, the smaller $\sum_{k=1}^m \squared{z_{T,k} - z_{C,k}}$ will be, the smaller the variance becomes, and the higher the power. If we further allow $n_R = 0$ (\textit{all} the subjects matched), then it's clear that $\sigsqDeltaYbar / \sigsqDbar > 1$ if and only if $\sum_{k=1}^m z_{T,k} z_{C,k} > 0$. Note that the approximation in the last expression is due to ignoring covariance terms which do not exist when conditioning on $n_T$ and $n_C$. 



\subsubsection{The Least Squares Test}\label{subsubsec:ols_test}

To construct a test when the response is linear in the covariates or when we wish to make linear adjustments, we extend the idea in the previous section where we combined an effect estimate from the matched pairs data and an effect estimate from the reservoir data to regression models. Consider the following model for the response differences among the matched pairs:

\beqn
D_k = \beta_{0,D} + \beta_{1,D} \Delta x_{1,k} + \ldots + \beta_{p,D} \Delta x_{p,k} +  \epsilonrv_{k,D}, \quad\quad \epsilonrv_{k,D} \iid \normnot{0}{\tausq_D}, \quad \quad k \in \braces{1, \ldots, m}.
\eeqn

The parameter of interest is the intercept, $\beta_{0,D}$, with OLS estimator $B_{0,D}$, the analogue of $\Dbar$ in the previous section. $\Delta x_{1,k}, \ldots, \Delta x_{p,k}$ are the differences between treatment and control within match $k$ for each of the $p$ covariates respectively. $\beta_{1,D}, \ldots, \beta_{p,D}$ are nuisance parameters that adjust for linear imbalances in the covariate differences not accounted for in the matching procedure. 

For the responses in the reservoir, consider the model:

\small
\beqn
Y_i = \beta_{T,R} \indic{T,R,i} + \beta_{0,R}  + \beta_{1,R} x_{1,i} + \ldots + \beta_{p,R} x_{p,i} +  \epsilonrv_{i,R}, \quad\quad \epsilonrv_{i,R} \iid \normnot{0}{\tausq_R}, \quad \quad i \in \braces{1, \ldots, n_R}.
\eeqn
\normalsize

The parameter of interest is the additive effect of the treatment, $\beta_{T,R}$, with OLS estimator $B_{T,R}$, the analague of $\YbarRTMinusYbarRC$ in the previous section. $\beta_{0,R}, \beta_{1,R}, \ldots, \beta_{p,R}$ are nuisance parameters that adjust for a mean offset and linear imbalances in the covariates. 

Using the parallel construction in equations \ref{eq:estimator} and \ref{eq:clt_true_var}, our modified OLS estimator has the form

\bneqn\label{eq:clt_true_var_linear}
\frac{B_{T,OLS} - \beta_0}{\se{B_{T,OLS}}} \approx \frac{\dfrac{S^2_{B_{T,R}} B_{0,D} + S^2_{B_{0,D}} B_{T,R}}{S^2_{B_{T,R}} + S^2_{B_{0,D}}} - \beta_0}{\sqrt{\dfrac{S^2_{B_{T,R}} S^2_{B_{0,D}}}{S^2_{B_{T,R}} + S^2_{B_{0,D}}}}}  ~\convd~ \stdnormnot
\eneqn

where $S^2_{B_{T,R}}$ is the sample variance of $B_{T,R}$ and $S^2_{B_{0,D}}$ is the sample variance of $B_{0,D}$.

\subsubsection{The Permutation Test}\label{subsubsec:permutation_test}

An application of Fisher's exact test is straightforward. For the matched pairs component of the data, we examine the $2^m$ configurations (each match can have T-C or C-T) to compute all $\dbar$'s. For the reservoir portion of the estimator, we condition on $\nrt$\footnote{This is known as the ``conditional'' exact test \citep{RosenbergerLachin2002}.} and examine every possible arrangement of the treatment vector to compute every $\ybarRTMinusybarRC$. For each arrangement, we also compute $\ssqDbar$ and $\ssqR$ to create values of the test statistic in equation \ref{eq:estimator}. In practice, the 2-sided $p$-value is approximated bv comparing the observed $b_T$ from the true sample data to Monte-Carlo samples from the space of all possible test statistics. A similar exact test is available using the modified OLS estimates.

\subsection{Properties of the Matching Algorithm}\label{subsec:theoretical}

We wish to gain insight about how $\lambda$ and $n$ affect $n_R$. Assume for now that we only have one covariate, $x$ (which may also be the largest principal component of a collection of covariates). Mahalanobis distance matches on standardized distance. For this illustration, assume we match if the two $x$'s sample quantiles are within $\lambda$ of each other. For example, the latest subject in the experiment had a sample quantile of 0.96, they would be matched to the closest subject in the reservoir with quantile between $0.91$ and 1 at $\lambda = 0.10$.

Consider dividing the unit interval into $K := 1/\lambda$ intervals of equal length. Two items in one interval qualify to be matched. Assume that $K$ is even (similar results follow for $K$ odd). Consider the Markov process that transitions after each pair of subjects. Let $s$ be the state that $2m$ of the $K$ cells are occupied for $s \in \braces{0, 1, \ldots, K/2}$. It is straightforward that $P_{i,j}$, the transition probability of pairs from state $i$ to $j$, satisfies:

\beqn
P_{s,j} = \begin{cases}
\frac{2s(2s - 1)}{K^2}, 								& j = s - 1, ~~ s \neq 0 \\
\frac{K(4s+1) - 8s^2}{K^2}, 							& j = s \\
\frac{K^2 - K(4s+1) + 2s(2s+1)}{K^2}, 				& j = s + 1, ~~ s \neq \overtwo{K}
\end{cases}
\eeqn

Note the inherent symmetry: $P_{s,j} = P_{K/2 - s, K/2 - j}$. Hence, the steady-state probabilities are symmetric about $s=K/4$. Therefore, the mean number of items in the reservoir goes to $K/2 = (2\lambda)^{-1}$ as $n$ grows arbitrarily large. For example, $\displaystyle \limitn \cexpe{N_R}{\lambda = 0.10} = 5$.

\section{Simulation Studies}\label{sec:simulations}

We demonstrate our method's performance by simulating in three scenarios: covariates affect the response non-linearly (the ``NL'' scenario), covariates affect the response linearly (the ``LI'' scenario) and covariates do not affect the response (the ``ZE'' scenario). These scenarios were simulated via the settings found in  table \ref{tab:model_coefficients}. In practice, we simulated many settings for the NL and LI scenarios with similar results.

\begin{table}[htp]
\centering
\begin{tabular}{c|l}
Scenario & $Y_i$ \\ \hline
NL & $ \betaT\indicTi + x_{1,i} + x_{2,i} + x_{1,i}^2 + x_{2,i}^2 +x_{1,i} x_{2,i} + \epsilonrv_i$ \\
LI & $ \betaT\indicTi + 2 x_{1,i} + 2 x_{2,i} + \epsilonrv_i$ \\
ZE & $\betaT\indicTi + \epsilonrv_i$ \\ \hline
\end{tabular}
\caption{The response models for the three scenarios proposed. The covariates were $X_{1,i} \iid \normnot{0}{1}$  and $X_{2,i} \iid \normnot{0}{1}$ and the errors were  $\epsilonrv_i \iid \normnot{0}{\sigsq_e}$. }
\label{tab:model_coefficients}
\end{table}

We set the treatment effect to be $\betaT = 1$. $n$ and $\lambda$ were varied over a grid found in table \ref{tab:simulation_grid}. We then used $\sigsq_e$ to modulate the resolution in our comparisons. We chose $\sigsq_e = 3$ to be a good balance because even at $n=200$ comparisons were clear. 

\begin{table}[htp]
\centering
\begin{tabular}{c|c}
Parameter & Simulated Values \\ \hline
$n$ & $\braces{50,100,200}$ \\
$\lambda$ & $\braces{0.05,0.075,0.10,0.20}$ \\ \hline
\end{tabular}
\caption{Parameters for simulation}
\label{tab:simulation_grid}
\end{table}

In choosing which competitor dynamic allocation methods to simulate against, we wanted to pick methods that are in use in sequential trials. According to \citet{Scott2002}, stratification is very popular and Efron's biased coin has been used in a few studies. Most popular is minimization which has been used in over 1,000 trials \citep{McEntegart2003} while $D_A$ optimality is not known to the authors to have been implemented to date and according to the simulations in \citet{Atkinson1999}, it does not perform dramatically better than minimization even though it rests on more solid theoretical grounds.

Thus, we choose to compare our method against complete randomization (CR), stratification,\footnote{Both $x_1$ and $x_2$ were stratified into three levels based on the 33.3\%ile and 66.6\%ile of the standard normal distributions. Thus, we create nine blocks. Within blocks we alternate T / C in order to coerce $n_T \approx n_C$ so no power is lost on allocation imbalance. We then ran OLS using the nine blocks as well as $x_1$ and $x_2$ as covariates.} Efron's biased coin design (BCD),\footnote{We use the bias parameter of $\alpha = 2/3$ which is Efron's ``personal favorite.''} and minimization.\footnote{We used the same blocks as stratification. For the ``$D$'' function, \citet{Begg1980} compared \citet{Pocock1975}'s range and variance methods using simulations and found the variance method performed slightly better, thus we implement variance as the ``$D$'' function and sum as the ``$G$'' function. We set $p=1$ for deterministic assignments in order to force $n_T \approx n_C$ so no power is lost on allocation imbalance.}

There are three scenarios (NL, LI, and ZE) and four competitors. Naturally, we want to gauge performance if we assumed the correct underlying model, but we also want to ensure we are robust if the model is misspecified. Therefore, we simulate each of these under the three model assumptions discussed in section \ref{subsec:estimator}. For the classic estimator, all competitors employed $\YbarT - \YbarC$; for the linear estimator, all competitors employed OLS; and for the exact test, all competitors employed the standard conditional permutation test. 

We hypothesize that in the case of no effects (the ZE scenario), we will slightly underperform against competitors under all three testing procedures because of the loss of power due to the lower effective sample size when analyzing paired differences. If the effects are linear, we hypothesize to do slightly worse against the OLS procedures due to lower sample size. Under all other scenarios and models, we expect to do better.

%


We simulated each scenario 1,000 times and for exact testing, we Monte-Carlo sampled 1,000 times within each simulation.\footnote{Original data and source code for reproduction can be found at \url{github.com/kapelner/sequential_matching_simulations}} The levels of $\lambda$ only minorly affect the comparisons against the four competitors. We empirically observed $\lambda = 10\%$ to be optimal, so we display those results.

\begin{figure}[htp]
\centering
\begin{subfigure}[b]{0.33\textwidth}
                \centering
                \includegraphics[width=\textwidth]{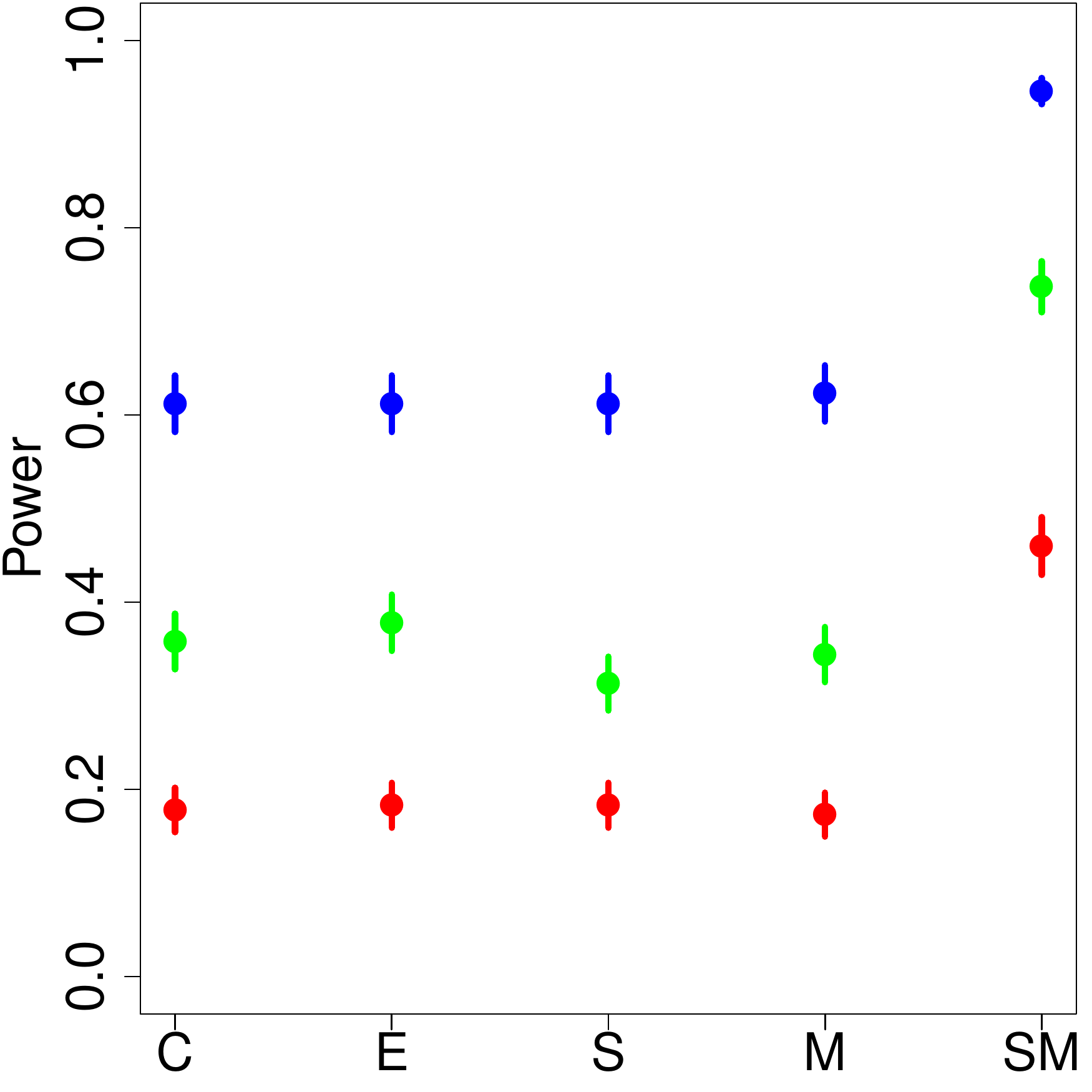}
                \caption{Scenario NL, Classic Test}
                \label{fig:power_results_model_I_classic}
        \end{subfigure}~~
\begin{subfigure}[b]{0.33\textwidth}
                \centering
                \includegraphics[width=\textwidth]{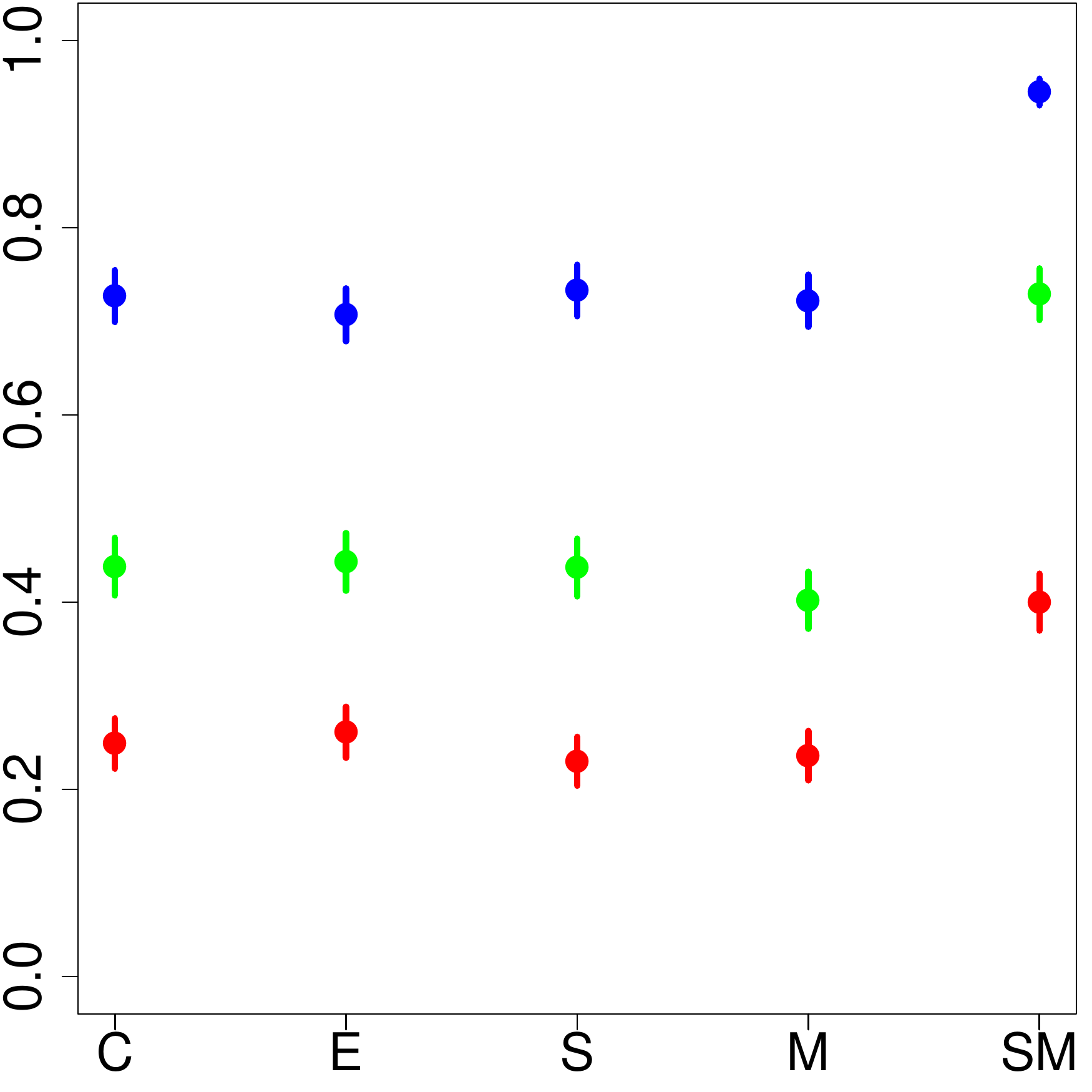}
                \caption{Scenario NL, Linear Test}
                \label{fig:power_results_model_I_linear}
        \end{subfigure}~~
\begin{subfigure}[b]{0.33\textwidth}
                \centering
                \includegraphics[width=\textwidth]{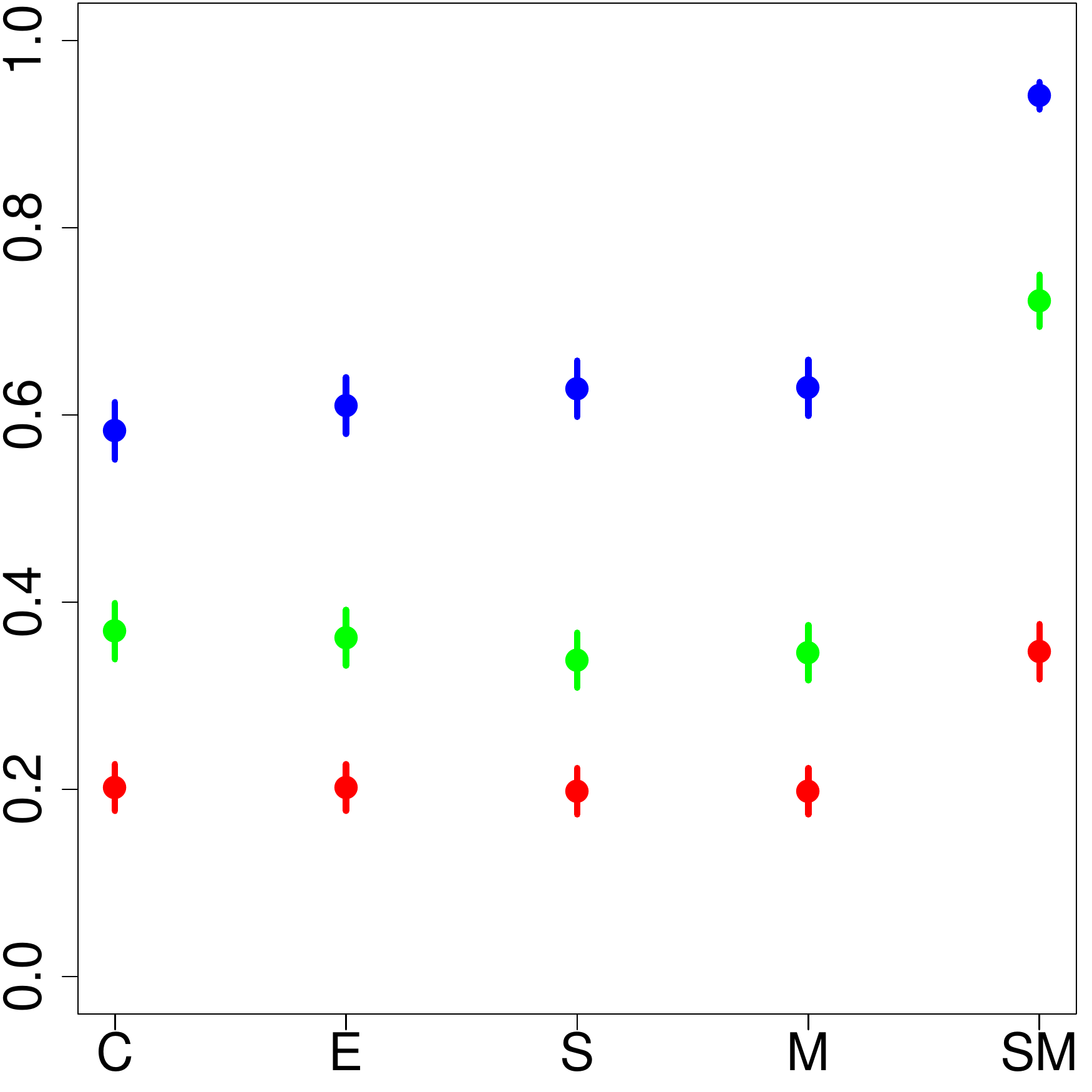}
                \caption{Scenario NL, Exact Test}
                \label{fig:power_results}
        \end{subfigure}\\ \vspace{0.3cm}
\begin{subfigure}[b]{0.33\textwidth}
                \centering
                \includegraphics[width=\textwidth]{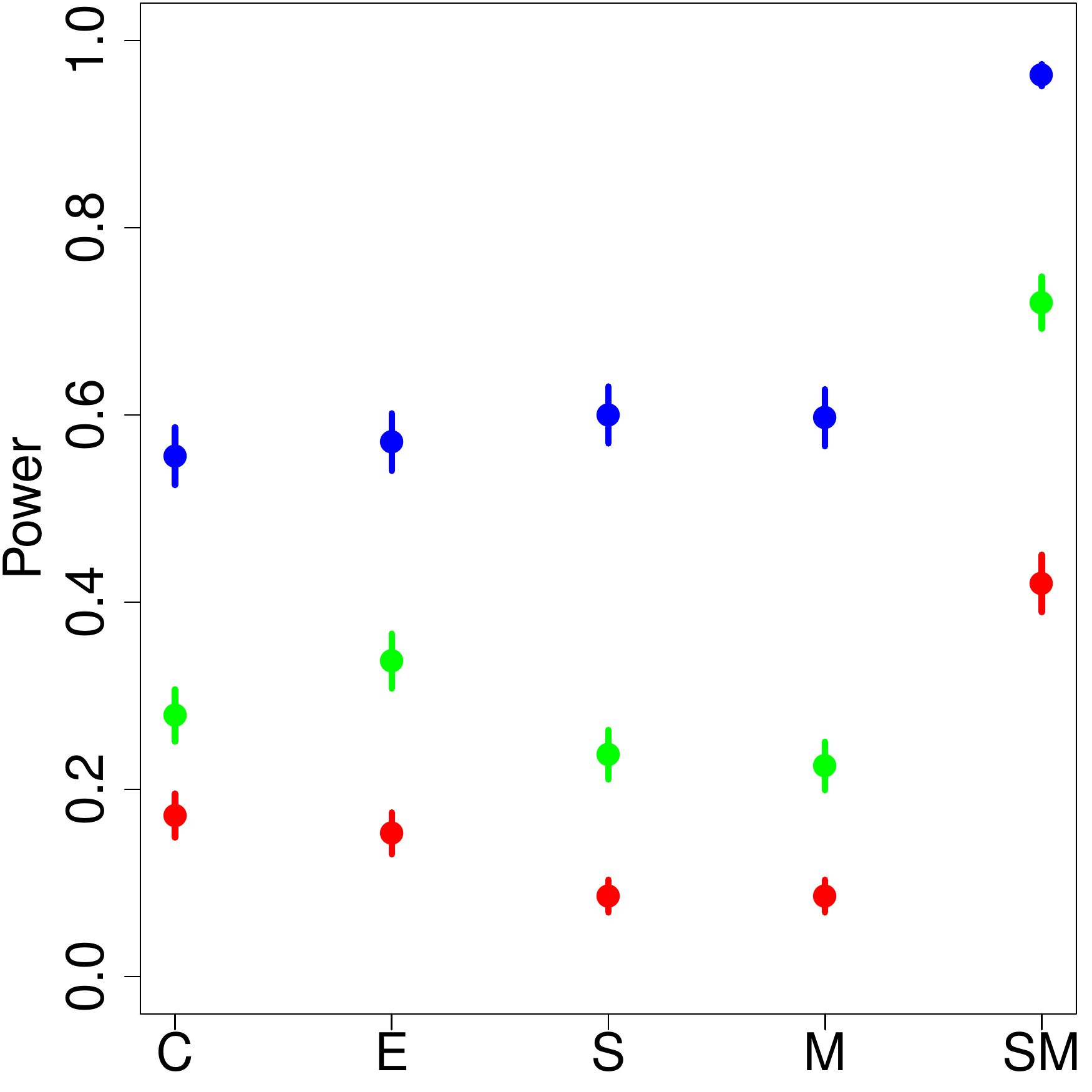}
                \caption{Scenario LI, Classic Test}
                \label{fig:power_results_model_II_classic}
        \end{subfigure}~~
\begin{subfigure}[b]{0.33\textwidth}
                \centering
                \includegraphics[width=\textwidth]{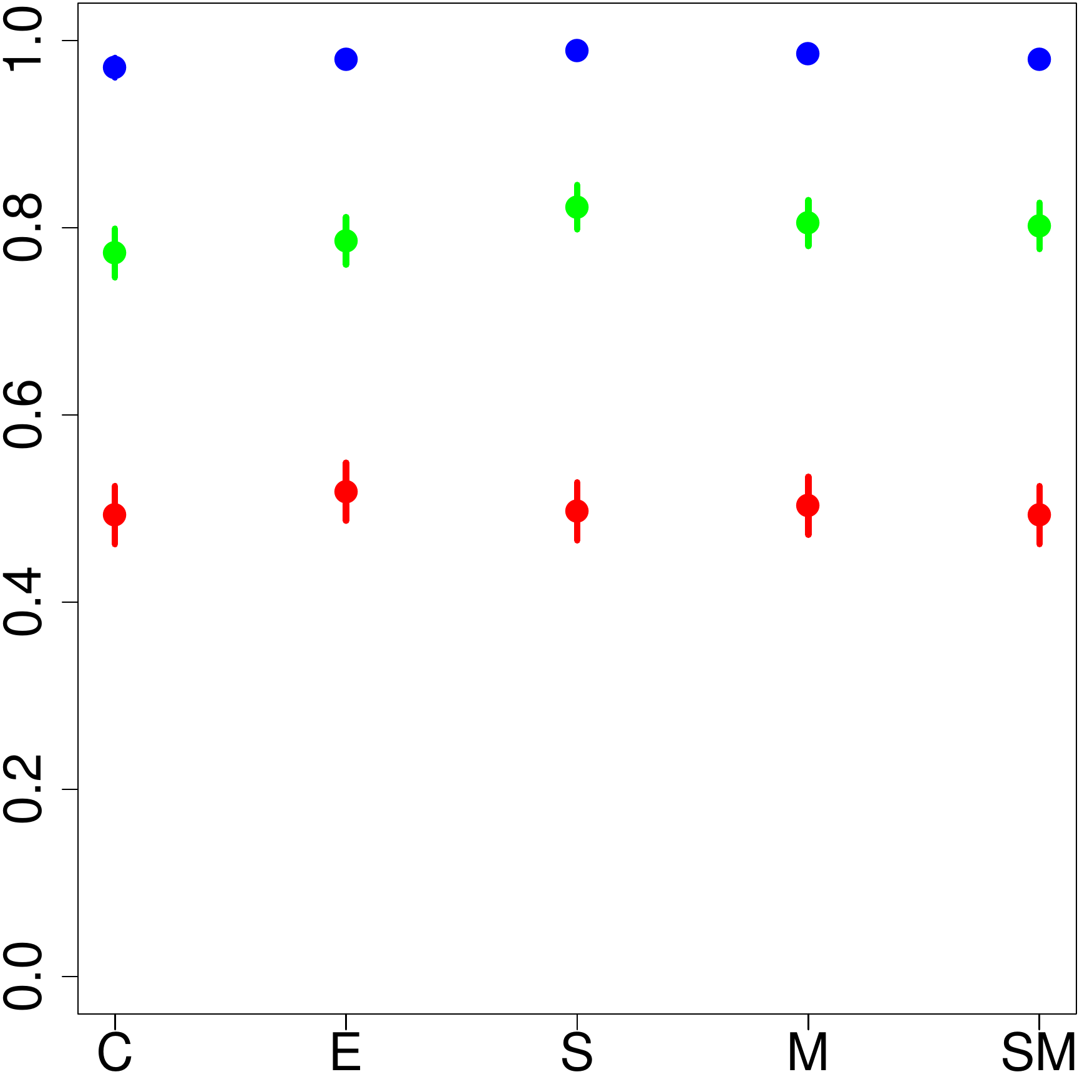}
                \caption{Model LI, Linear Test}
                \label{fig:power_results_model_II_classic}
        \end{subfigure}~~
\begin{subfigure}[b]{0.33\textwidth}
                \centering
                \includegraphics[width=\textwidth]{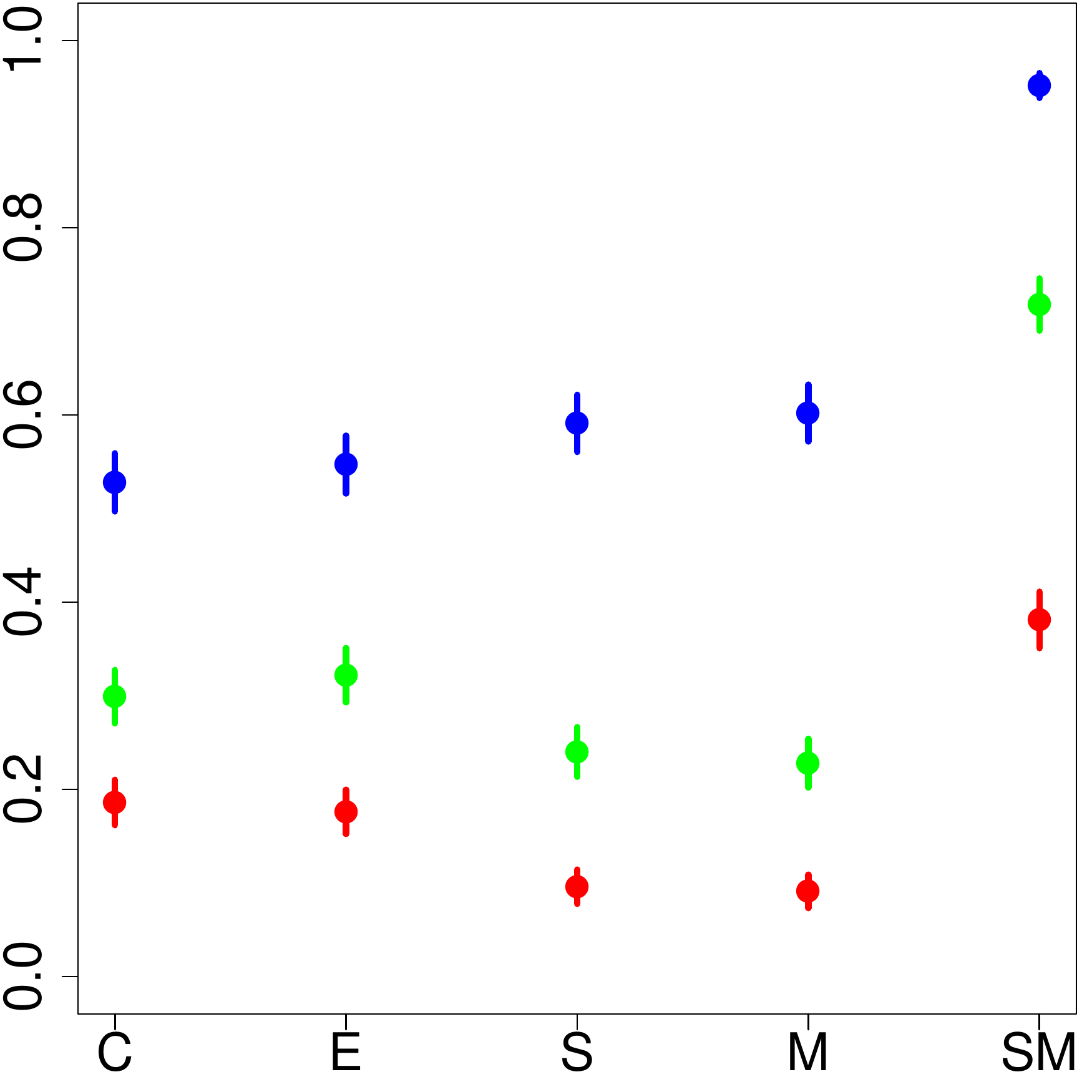}
                \caption{Scenario LI, Exact Test}
                \label{fig:power_results_model_II_classic}
        \end{subfigure}\\ \vspace{0.3cm}
\begin{subfigure}[b]{0.33\textwidth}
                \centering
                \includegraphics[width=\textwidth]{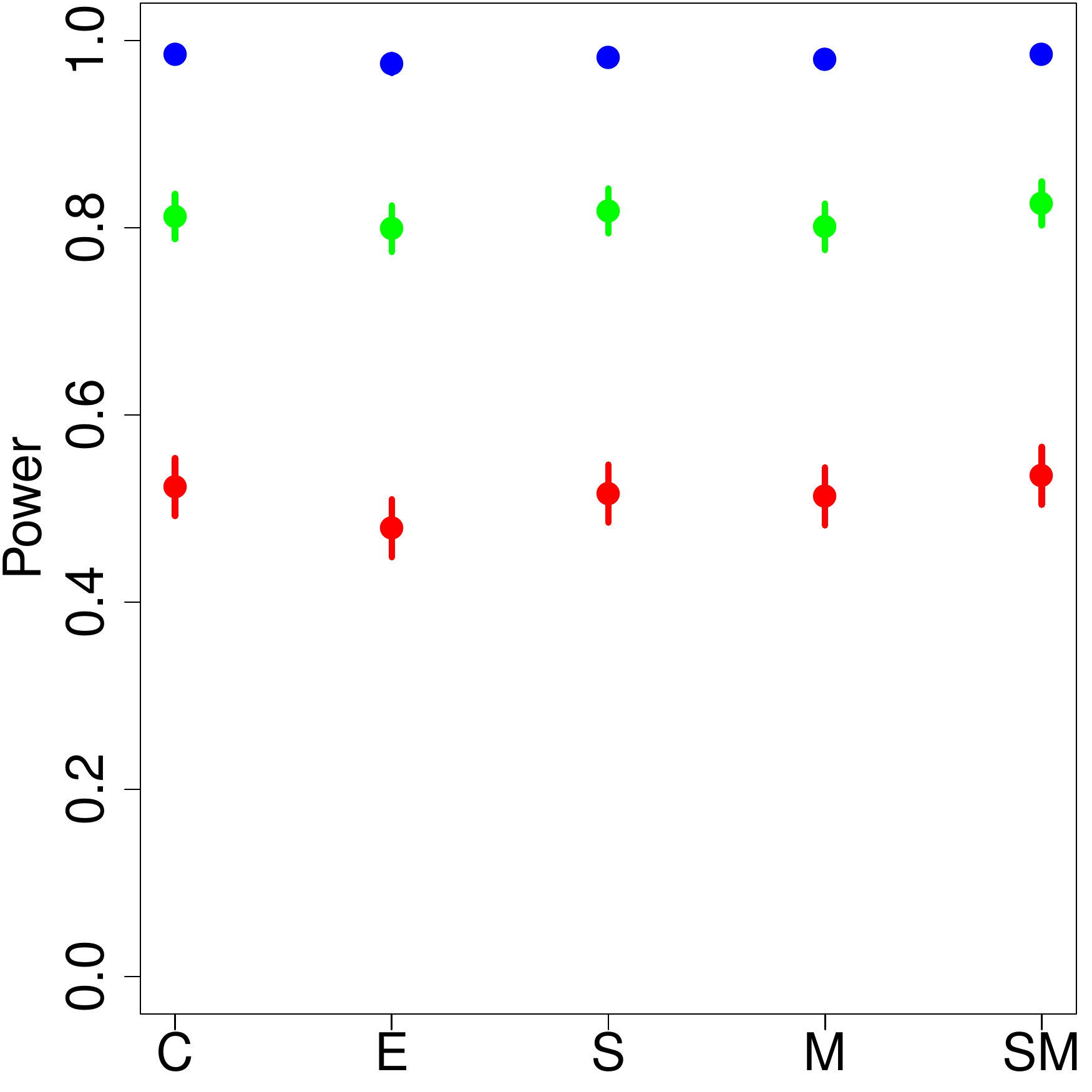}
                \caption{Scenario ZE, Classic Test}
                \label{fig:power_results_model_II_classic}
        \end{subfigure}~~
\begin{subfigure}[b]{0.33\textwidth}
                \centering
                \includegraphics[width=\textwidth]{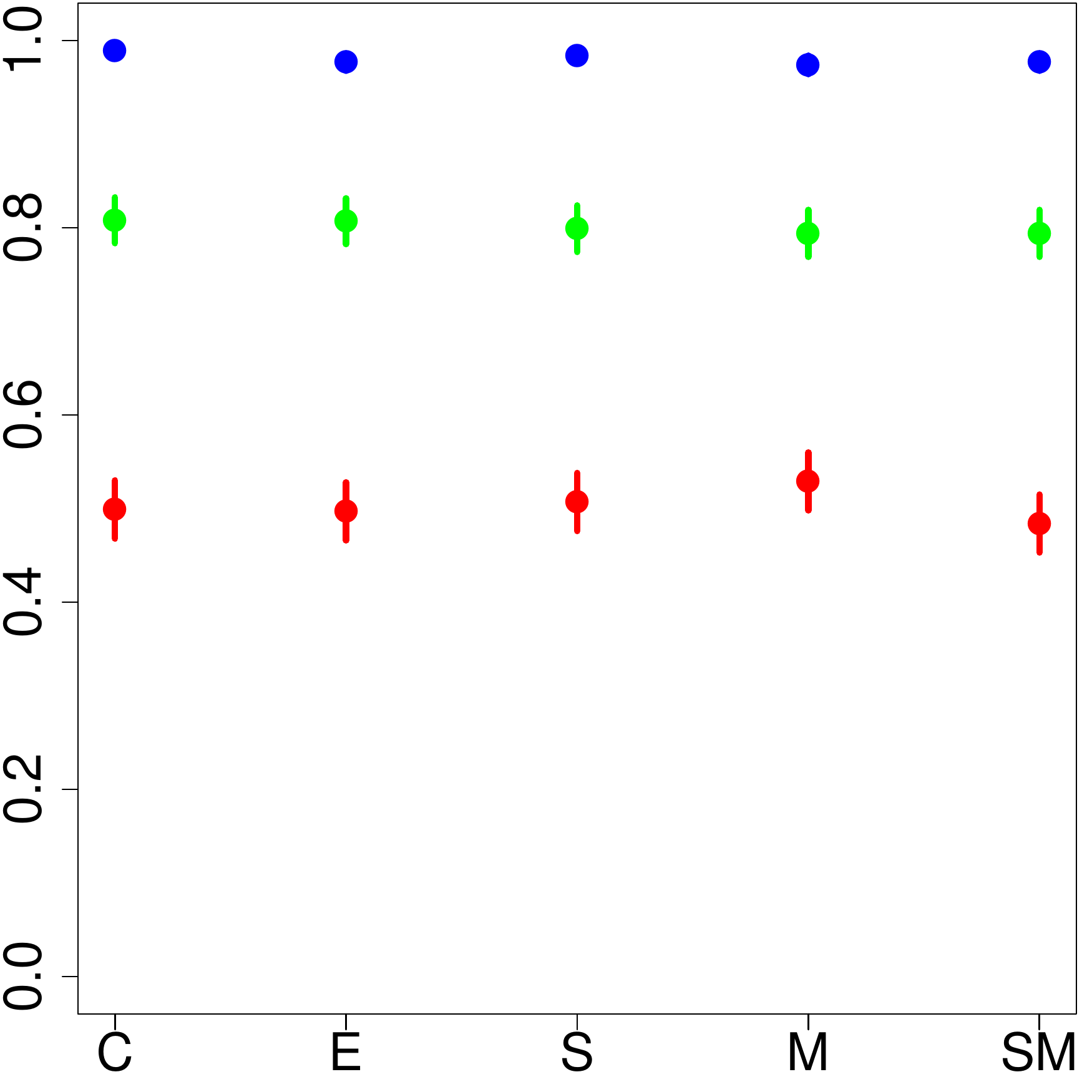}
                \caption{Scenario ZE, Linear Test}
                \label{fig:power_results_model_II_classic}
        \end{subfigure}~~
\begin{subfigure}[b]{0.33\textwidth}
                \centering
                \includegraphics[width=\textwidth]{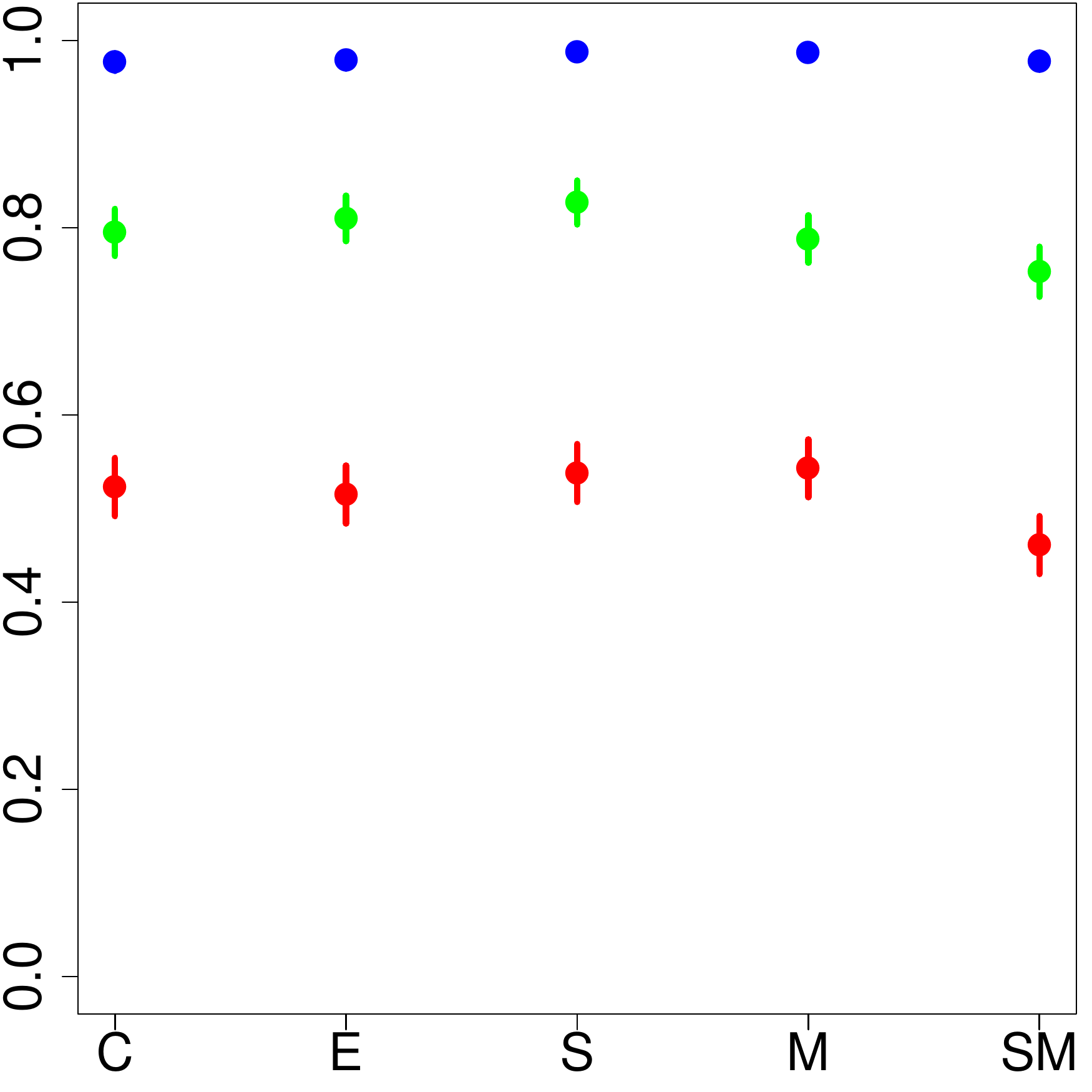}
                \caption{Scenario ZE, Exact Test}
                \label{fig:power_results_model_II_classic}
        \end{subfigure}\\
\caption{Power illustrated for the three scenarios by the three testing procedures, all five allocation methods (C: Complete Randomization, E: Efron's BCD, S: Stratification, M: Minimization, SM: our sequential matching algorithm) and sample sizes (\inred{red~} illustrates results for $n=50$, \ingreen{green~} for $n=100$ and \inblue{blue~} for $n=200$). Plotted points represent the sample proportion of null hypothesis rejections in 1,000 simulations and segments represent 95\% confidence intervals. Matching parameter $\lambda = 10\%$.}
\label{fig:power_results}
\end{figure}

\begin{table}[htp]
\centering
\begin{tabular}{cc|c|cc|cc|cc}
\multicolumn{3}{c}{} & \multicolumn{6}{c}{\underline{\smash{Sample Relative Efficiency Over Competitors}}}            \\
& Allocation &      & \multicolumn{2}{|c|}{Scenario NL}  & \multicolumn{2}{|c|}{Scenario LI}  & \multicolumn{2}{|c}{Scenario ZE}           \\
$n$ & Method                           & Balance & Classic        & Linear        & Classic       & Linear       & Classic      & Linear \\ \hline
\multirow{5}{*}{50}  & CR                & 0.816    & \ingreen{1.954}  & \ingreen{1.721} & \ingreen{2.572} & \inred{0.756} & 0.983 & 0.898 \\
                                 & Efron's BCD   & 0.815    & \ingreen{2.048} & \ingreen{1.576} & \ingreen{2.344} & \inred{0.752} & 0.915 & \inred{0.805}\\
                                 & Stratification  & 0.423   & \ingreen{1.484} & \ingreen{1.236} & \ingreen{1.198} & \inred{0.613} & 0.930 & \inred{0.834}\\
                                 & Minimization  & 0.395    & \ingreen{1.694} & \ingreen{1.426} & \ingreen{1.141} & \inred{0.692} & 0.961 & \inred{0.842}\\
                                 & Seq. Matching & \textbf{0.587} & --- &  --- &  ---  & ---  &---  & ---  \\ \hline
\multirow{5}{*}{100} & CR               & 0.798	 & \ingreen{2.545} & \ingreen{1.721} & \ingreen{2.748} & 1.020 & 0.879 & 0.903 \\
                                 & Efron's BCD   & 0.808    & \ingreen{2.687} & \ingreen{1.576} & \ingreen{2.904} & 0.956 & 1.001 & 0.878\\
                                 & Stratification  & 0.390    & \ingreen{1.893} & \ingreen{1.236} & \ingreen{1.234} & 0.875 & 0.887 & 0.911\\
                                 & Minimization  & 0.369    & \ingreen{2.111} & \ingreen{1.426} & \ingreen{1.220} & 0.907 & 0.970 & 0.881\\
                                 & Seq. Matching & \textbf{0.497} & ---  & ---   &  ---  & ---  & ---  & ---  \\ \hline  
\multirow{5}{*}{200} & CR               & 0.823    & \ingreen{2.649} & \ingreen{1.942} & \ingreen{2.947} & 1.029 & 0.980 & \inred{0.852}		\\
                                 & Efron's BCD   & 0.812    & \ingreen{2.541} & \ingreen{1.948} & \ingreen{3.155} & 1.003 & 1.033 & 0.970\\
                                 & Stratification  & 0.379    & \ingreen{1.585} & \ingreen{1.451} & \ingreen{1.475} & 0.905 & 0.951 & 0.913\\
                                 & Minimization  & 0.369    & \ingreen{1.821} & \ingreen{1.750} & \ingreen{1.159} & 0.892 & 0.977 & 0.995\\
                                 & Seq. Matching & \textbf{0.419} & ---   & ---  & ---  & ---  & --- & ---   \\ \hline
\end{tabular}
\caption{Balance results and relative sample efficiency results of sequential matching versus competitors by scenario and testing procedure. Efficiencies in \inred{red~} indicate our algorithm performed worse than a competitor via an F-test with 1\% significance level and efficiencies in \ingreen{green~} indicate our algorithm performed better. Note that we did \textit{not} adjust for multiple comparisons. Balance results are averages over all scenarios and two model assumptions (6,000 simulations). Exact tests are not shown because they do not admit a standard error calculation.}
\label{tab:balance_and_efficiency_results}
\end{table}

Our main metric for comparison is power, the proportion of the times the null was rejected under the Type I error rate of $\alpha = 5\%$. We also record standard error of the estimate (when the estimator was parametric) as well as balance.\footnote{By ``balance'' we mean the maximum standardized difference in the averages of covariates between treatment and control samples: $\max_{j \in \braces{1,2}} \braces{(\xbar_{j,T} - \xbar_{j,C}) / s_j}$.} Results for power against the null of no treatment effect are illustrated in figure \ref{fig:power_results} and results for balance and relative efficiency vis-a-vis other methods are found in table \ref{tab:balance_and_efficiency_results}.

In the NL scenario, our sequential matching procedure dominates competitors in power and efficiency, sometimes doubling power and nearly tripling efficiency. Even at $n=200$, there are still gains. Regression adjustment helps the competitors, but it cannot adjust for the non-linear portion of the quadratic terms and interaction term; they will appear as higher noise.

In the LI scenario, sequential matching dominates competitors in the classic and the exact test because the competitors do not use the covariate information. In the linear assumption, sequential matching performs similarly in power but has a lower efficiency at $n=50$. This loss is due to a lower effective sample size due to using matched pairs. The loss does not continue at $n > 50$. This can be explained that there are benefits to matching even when employing regression adjustment. As \citet{Greevy2004} explain, better balance reduces collinearity resulting in a smaller standard error for the estimate. Balance is improved over competitors that do not allocate based on the covariates and this better balance implies higher power and efficiency. Parenthetically, we note that as $n$ increases, it appears as if balance is approaching levels observed in both stratification and minimization. This is expected and is an added bonus of our procedure.

In the ZE scenario, our approach is most severely impacted by lower effective sample size. However, power is not as low as expected. Efficiency seems to be lost for all simulated $n$ but most significantly when $n=50$.

All in all, sequential matching performs well in the scenario of the response being non-linear in the covariates which is the most realistic case in practice. If the covariate model is truly linear, we only do worse when OLS is employed but our inefficiency is only for small sample sizes. In the case when covariates do not matter at all, we do worse for low sample size, but about equal when $n \geq 100$.

All in all, sequential matching shines in the case of non-linear covariate models which is the most realistic case in practice. If the covariate model is truly linear, sequential matching does worse when OLS is employed but our relative inefficiency is only observed for small sample sizes. In the case when covariates do not matter at all, we begin to perform about equally with competitors when $n \geq 100$. This is an important result in practice because investigators sometimes choose useless covariates which do not affect the outcome measure.

A possible criticism of the high power achieved is we assume our $n$ was large enough for the estimators in equations \ref{eq:clt_true_var} and \ref{eq:clt_true_var_linear} to converge. To assuage this concern, we simulated the size of the tests in table \ref{tab:size_of_tests}. For the classic estimator, even at $n=50$, the size is about 7-8\% and by $n=100$ it's close to the Type I error rate of 5\%. For the OLS estimator, convergence is a tad slower. In unshown simulations, we have observed that the modified classic estimator can be approximated by a $T_{m-1}$ distribution which can be used to compute a more conservative significance level. Even under this conservative approximation, the simulated power in figure \ref{fig:power_results} do not change dramatically (unshown). Other anomalies observed in this table are discussed in section \ref{sec:discussion}.

\begin{table}[htp]
\centering
\small
\begin{tabular}{cc|ccc|ccc|ccc}
 & Allocation & \multicolumn{3}{|c|}{Scenario NL} & \multicolumn{3}{|c|}{Scenario LI} & \multicolumn{3}{|c}{Scenario ZE} \\
$n$ & Method                           & Classic & Linear & Exact & Classic & Linear & Exact & Classic & Linear & Exact \\ \hline
\multirow{5}{*}{50}  & CR                 & 0.039 & 0.038 & 0.058 & 0.041 & 0.057 & 0.051 & 0.048 & 0.060 & 0.042 \\
                                 & Efron's BCD    & 0.056 & 0.049 & 0.056 & 0.046 & 0.051 & 0.049 & 0.049 & 0.049 & 0.052 \\
                                 & Stratification   & \inred{0.019} & \indarkred{0.026}  & \inred{0.016} & \inred{0.005} & \inred{0.010} & 0.048 & 0.048 & 0.051 & 0.048 \\
                                 & Minimization   & \indarkred{0.036} & 0.039 & 0.043 & \inred{0.007} & \inred{0.004} & 0.061 & 0.049 & 0.049 & \indarkred{0.067} \\
                                 & Seq. Matching & \inred{0.075} & \inred{0.078} & 0.051 & \indarkred{0.071} & 0.051 & \inred{0.080} & \indarkred{0.073} & \indarkred{0.068} & 0.044 \\ \hline
\multirow{5}{*}{100} & CR                & 0.041 & 0.039 & \indarkred{0.035} & 0.043 & 0.054 & 0.048 & 0.049 & \indarkred{0.066} & 0.052 \\
                                 & Efron's BCD    & 0.059  & 0.059 & 0.062 & 0.062 & 0.063 & 0.052 & \indarkred{0.035} & 0.040 & 0.055 \\
                                 & Stratification   & \indarkred{0.023} & 0.040& \inred{0.023} & \inred{0.006} & 0.045 & \inred{0.002} & 0.056 & 0.038 & 0.051 \\
                                 & Minimization   & \inred{0.016} & \indarkred{0.036} & 0.043 & \inred{0.003} & 0.049 & \inred{0.008} & 0.046 & 0.041 & 0.040 \\
                                 & Seq. Matching & \indarkred{0.066} & \inred{0.076} & 0.047 & 0.058& 0.058 & 0.058 & 0.063 & \indarkred{0.071} & 0.044 \\ \hline
\multirow{5}{*}{200} & CR                & 0.042 & 0.042 & 0.042 & 0.047 & 0.048 & 0.056 & 0.054 & 0.052 & 0.045 \\
                                 & Efron's BCD    & 0.056 & 0.046& 0.041 & 0.047 & 0.044 & 0.055 & 0.054 & 0.053 & 0.060 \\
                                 & Stratification   & \inred{0.020} & \indarkred{0.033} & \inred{0.023} & \inred{0.001} & 0.050 & \inred{0.005} & 0.045 & 0.042 & 0.050 \\
                                 & Minimization   & \inred{0.025} & 0.044 & 0.051 & \inred{0.001} & 0.045 & \inred{0.001} & 0.059 & \indarkred{0.064} & 0.056 \\
                                 & Seq. Matching & 0.050 & 0.062 & 0.051 & 0.046 & 0.059 & 0.056 & 0.048 & \indarkred{0.065} & \indarkred{0.065} \\ \hline
\end{tabular}
\caption{Simulated size of tests for all scenarios, competitors, and all tests at $\lambda = 10\%$. Numbers in \inred{red~} indicate they are different from the purported 5\% size at a Bonferroni-corrected significance level (135 comparisons). Numbers in \indarkred{orange~} indicate they are different from the purported 5\% size without Bonferroni correction.}
\label{tab:size_of_tests}
\end{table}

\section{Demonstration Using Real Data}\label{sec:real_data}

We now examine sequential experiment data from two real applications: one from a behavioral study on the Internet and one from a double-blind drug trial. We simulate the subjects being dynamically allocated using the sequential matching procedure by first assuming the entering subjects do not exhibit any time trend; this will allow us to permute their order. During the iterative procedure, all subjects assigned to the reservoir keep whichever assignment they had during the experiment. During matching, if the subject happened to have been assigned the treatment which SM allocates, they are kept in the subject pool; if not, they are discarded (this is illustrated in figure \ref{fig:historical_data_runs}). Thus, during our simulations, we result in a \textit{subset} of the data we began with. Note that we only show results for the classic estimator versus the modified estimator in equation \ref{eq:clt_true_var}, not the OLS modified estimator whose results we suspect to be similar.

\begin{figure}[htp]
\centering
\small
\begin{BVerbatim}[baseline=c]
nsim: 1  .............o.xx....o....xxx.ooxx.o.xxooxooxx.xo.  (37)
nsim: 2  ................xxx..x.xxoo.xxx.o.o.oo..ox.xxxxx.o  (35)
nsim: 3  ................x.x..oo.xxxo.x.xxxxxxxo.x.ox.oox.o  (34)
nsim: 4  ...........o...o...ox.o.o.xo.ox.o....xxooox.oox.xo  (42)
nsim: 5  ................x.xoo.o.o..xxoxx.o.x.x.oo.o.xoxx..  (39)
nsim: 6  ................x..xx.x..xx.oo.xoxoxxx.ooxo.o.xxxo  (35)
nsim: 7  .............x.o..oo..oxx.x..o.oxo.xxox.o.xoxxxoxo  (37)
nsim: 8  ............xx.oxoxxo..x.xxx.x..oxoo.o.xx.x..xox.o  (34)
nsim: 9  ..............o...xxoo.xo.ooo.....o.ooxox.oxx..o.x  (42)
\end{BVerbatim}
\caption{Running an $n=50$ subset of historical data through the sequential procedure. The dots represent a subject being placed into the reservoir. The ``o'' signifies that the subject was matched and that their treatment allocation was \textit{opposite} of their matching partner. The ``x'' signifies that the subject was matched but their treatment allocation was the \textit{same} as their matching partner, resulting in the subject being discarded. The number in parenthesis at the end of the line is the sample size retained of the purported 50.}
\label{fig:historical_data_runs}
\end{figure}

\subsection{Behavioral Experiment}\label{subsec:behavior_data}

\citet{Kapelner2010} ran experiments using the Amazon Mechanical Turk platform, a global outsourcing website for small one-off tasks that can be completed anonymously on the Internet. They focused on measuring subjects' stated preference for a beer price when the beer came from different purchasing locations (an online replication of \citealp{Thaler1985}'s demonstration of the ``framing effect,'' a cognitive bias). The treatment involved subtle text manipulations: the same beer came from either a \textit{fancy resort} or a \textit{run-down grocery store}. In their control wing ($n=168$), no tricks were employed to ensure the subjects were paying attention to the text. Thus, in this wing, the subtle text manipulation did not seem to affect the subjects' stated beer prices. The effect may have been real, but the data was either too noisy or there was insufficient sample size to find it. We demonstrate here that if our sequential matching procedure was employed, the effect estimator would have been more efficient.

For matching, we first used most of the covariates found in the original dataset: age, gender,  level of earnings, number of weekly hours spent doing one-off tasks, level of multitasking when performing tasks, stated motivation level, passing the ``instructional manipulation check''  \citep{Oppenheimer2009} and a survey gauging the subject's ``need for cognition'' \citep{Cacioppo1982}. 

We note that $R^2$ under OLS was about 18.7\%. We then run two off-the-shelf machine learning algorithms that are designed to find interactions and non-linearities in the response function. The in-sample pseudo-$R^2$ using \citet{Chipman2010}'s Bayesian Additive Regressive Trees  (BART) was 42.4\% and \citet{Breiman2001}'s Random Forests (RF) was 70.4\%. Although this is not a formal test, it is pretty compelling evidence that the covariates do not combine strictly linearly to inform beer price. Thus, as demonstrated in figure \ref{fig:power_results_model_I_classic} and column 4 of table \ref{tab:balance_and_efficiency_results}, our method should be more powerful and more efficient than using previous dynamic allocation strategies with a classic estimator. The results for 200 simulations at $\lambda = 0.10$ are shown in table \ref{tab:kapcha_all_covs}. Many of the covariates are binary. Thus, the variance-covariance matrix was \textit{not} invertible in line 5 of algorithm \ref{alg:matching} for many of the early iterations, so we used the Moore-Penrose generalized inverse instead.

\begin{table}[htp]
\centering
\begin{tabular}{r|ccc}
purported $n$ & actual $n$ (average) & average efficiency & approximate sample size reduction \\ \hline
50 & 37.8 & 1.84 & 45.7\% \\
100 & 71.9 & 1.23 & 16.9\% \\
168 (all) & 116.1 & 1.06 & 5.4\%
\end{tabular}
\caption{Results for 200 simulations over many values of $n$ and $\lambda = 0.10$ in the case where most of the covariates are matched on.}
\label{tab:kapcha_all_covs}
\end{table}

We now match on four selected covariates that come out most significant in an OLS regression on the full dataset: age, level of earnings, level of multitasking when performing tasks and one question from the survey gauging the subject's ``need for cognition.'' The results are found in table \ref{tab:kapcha_restricted_covs}. Note that the efficiencies are higher and do not drop off as quickly when $n$ increases. Thus, matching on \textit{relevant} covariates yields a performance enhancement in our procedure.

\begin{table}[htp]
\centering
\begin{tabular}{r|ccc}
purported $n$ & actual $n$ (average) & average efficiency & approximate sample size reduction \\ \hline
50 & 34.9 & 2.01 & 50.1\% \\
100 & 67.8 & 1.60 & 37.3\% \\
168 (all) & 112.1 & 1.57 & 36.3\%
\end{tabular}
\caption{Results for 200 simulations over many values of $n$ and $\lambda = 0.10$ in the case where four cherry-picked covariates are matched on. OLS has an $R^2 = 20.8\%$, BART, 33.1\% and RF, 26.5\%. }
\label{tab:kapcha_restricted_covs}
\end{table}

\subsection{Clinical Trial}\label{subsec:behavior_data}

We use data from \citet{Foster2010}, a twelve-week, multicenter, double-blind, placebo-controlled clinical trial studying whether amitriptyline, an anti-depressant drug, can effectively treat painful bladder syndrome. The study measured many outcomes, including change in pain after 12 weeks (difference in Likert scale scores). The confidence interval for the ATE between pill and placebo for this outcome measure was $\bracks{-1.00, 0.30}$ with a significance level of $\pval = 0.29$ (table 2, row 1, page 1856). Most likely the effect is real but there wasn't enough power to detect it either via a low sample size or a high error variance.

Once again, for matching, we first used most of the covariates found in the original dataset: age, gender, race (white / hispanic), level of education, level of employment, living with a partner, presence of sexually transmitted diseases and urinary tract infection, as well as baseline measures of pain, urination frequency and urgency, quality of life, anxiety and depression as well as syndrome symptom levels.  We note that $R^2$ was about 25.2\% under OLS, 42.6\% under BART, and 82.4\% under RF which is once again compelling evidence that the covariates do not combine strictly linearly to inform the subject's week 12 pain measure. By figure \ref{fig:power_results_model_I_classic} and column 4 of table \ref{tab:balance_and_efficiency_results}, our procedure should be more powerful and more efficient at estimating the ATE. The results for 200 simulations are shown in table \ref{tab:clin_trial_all_covs}. The covariates again include many binaries, so we used the Moore-Penrose generalized inverse for the inverse variance-covariance matrix calculation.

\begin{table}[htp]
\centering
\begin{tabular}{r|ccc}
purported $n$ & actual $n$ (average) & average efficiency & approximate sample size reduction \\ \hline
50 & 38.9 & 1.30 & 23.0\% \\
100 & 75.2 & 1.10 & 9.2\% \\
150 & 111.3 & 1.05 & 4.9\% \\
224 (all) & 165.5 & 1.07 & 6.7\%
\end{tabular}
\caption{Results for 200 simulations of the sequential matching procedure over many values of $n$ and $\lambda = 0.10$ in the case where most of the covariates are matched.}
\label{tab:clin_trial_all_covs}
\end{table}

Once again, we now match on the top four covariates which are the most significant in an OLS of the full dataset: living with a partner and baseline pain, frequency and syndrome symptom levels. The results are found in table \ref{tab:clin_trial_restricted_covs}. Note again that the efficiencies are higher when comparing to matching on all covariates and they do not suffer the steep drop off as $n$ increases. 

\begin{table}[htp]
\centering
\begin{tabular}{r|ccc}
purported $n$ & actual $n$ (average) & average efficiency & approximate sample size reduction \\ \hline
50 & 38.0 & 1.27 & 21.2\% \\
100 & 72.9 & 1.23 & 18.8\% \\
150 & 108.6 & 1.15 & 13.2\% \\
224 (all) & 160.3 & 1.13 & 11.3\%
\end{tabular}
\caption{Results for 200 simulations of the sequential matching procedure over many values of $n$ and $\lambda = 0.10$ in the case where four cherry-picked covariates are matched on. OLS has an $R^2 = 18.9\%$, BART, 35.1\% and RF, 45.0\%. }
\label{tab:clin_trial_restricted_covs}
\end{table}

\section{Discussion}\label{sec:discussion}

Estimation in sequential experiments can have higher power and efficiency if the covariate information is leveraged. We present a dynamic allocation of treatment and control that matches subjects on-the-fly via a novel algorithm and present modified estimators of classic approaches: average difference, linear regression, and permutation testing. We simulate under different scenarios and illustrate higher power in scenarios where competing methods cannot make proper use of covariate information. We underperform only in the case of low sample size when the covariate model is linear or non-existent. In simulations with real data, we find the efficiency of our method increases as the covariate function becomes more important. In the two real data sets from a clinical trial and an online behavioral study, our method is more efficient over complete randomization. This is most likely due to the fact that real-world response functions combine covariates non-linearly, and this is when our procedure is most advantageous. 

We note that ``analysis assumptions may be compromised due to the `pseudo'-random allocation'' \citep{Scott2002} and would like to address this criticism which can be made about our procedure.  Note that in table \ref{tab:size_of_tests}, the size of the tests under stratification and minimization are less than 5\%. One should not use the classic estimator in these cases because one implicitly tried to balance on the covariates, but then did not include the covariates in the model (as seen by the terrible sizes using the classic model in scenario NL using the classic model in scenario LI in table \ref{tab:size_of_tests}). \citet{Simon1979} and \citet{Senn2000} have very good discussions about this issue and recommend using regression adjustment (as seen in the competitors in scenario LI, model linear in the table \ref{tab:size_of_tests} for $n>50$). As for exact testing under dynamic allocation, \citet{Kalish1987} and many others warn that permutation distributions in stratification and minimization are incorrect unless the investigator permuted the treatment allocation \textit{according to its structure}, that is, according to how the allocation was determined by the covariates. This is not straightforward in practice for stratification and even more complicated for minimization.

In contrast, the sizes of the test for our apporach seem to be correct in table \ref{tab:size_of_tests} (barring the convergence of our estimator which we talked about at the end of section \ref{sec:simulations}). Usage of the classic estimator even for our method seems ill-advised for the same reasons that it is not recommended under stratification and minimization. However, using linear regression on the covariates is also ill-advised when the model is non-linear or otherwise does not satisfy the OLS model assumptions \citep{Freedman2008}. However, \citet{Rubin1973, Rubin1979} finds that covariance adjustment of matched pair differences is very robust to model misspecification.

Thus, since our permutation test performs well in the non-linear case and can avoid many of the above issues, we recommend the  permutation test in practice. We permuted according to the structure of our dynamic allocation via matching, thus our permutation tests are valid (this is also a reason not to employ anything but complete randomization in our reservoir). 

\subsection{Further Developments}

We view this contribution as a step forward in covariate-adaptive randomization in sequential experiments but it is far from complete. We begin with extensions that can be immediately implemented.

Although we assume fixed $n$ in our construction, it is relatively straighforward to adapt to a fully or group sequential design whose methods can be found in \citet{Jennison2000}. It would be hard to tabulate values when our estimator has unknown convergence properties, thus it would probably have to be done by waiting until the estimator most likely converged, and then using standard sequential analysis software. 



Fruitful areas of further work would be extensions to $k>2$ treatments which can be done by making matches of size $k$. We believe our methods can apply beyond continuous responses to binary, ordinal, or count responses but this will involve some adjustment of the estimators.  Fixed design is a simplifying assumption, but unrealistic with real-world covariates, begging a random-X robust implementation. The estimators found in \citet{Pitkin2013} can also be plugged into equations \ref{eq:clt_true_var} and \ref{eq:clt_true_var_linear}. We also believe there may be more clever estimators that can be constructed using the data in the matched pairs and reservoirs. We previously tried random effects models without success, but there may be others. Additionally, it would be straightforward to introduce a biased coin design to the matching algorithm to avoid possible tampering due to the partially deterministic allocation. 

We feel that the most significant improvement would be better matching. Mahalanobis distance is logical, but prone to strange behavior with departures from the normality assumption. Another natural extension is to boostrap the distribution of the nominal metric in equation \ref{eq:mahalanobis} as to not rely on probabilities from the scaled $F$ distribution. Also, practitioners may want to weight the variables in the matching as well as force some variables to always match (these ideas and others are discussed in \citealt[][chapter 8]{Rosenbaum2010}). Additionally, in the observational study literature, matching is elaborately engineered to improve balance across entire groups of observations \citep{Zubizarreta2012}. It is possible some of these methods can be applied to better matching for single pairs.

Of course, our procedure also suffers from the central issue of matching: selecting the variables to match on. A poor choice makes a big difference as evidenced by the simulations on historic RCT data (table \ref{tab:kapcha_all_covs} vs. \ref{tab:kapcha_restricted_covs} and table \ref{tab:clin_trial_all_covs} vs. \ref{tab:clin_trial_restricted_covs}). There may be a way to \textit{iteratively} match on covariates that are found to be important, so the set of perceived important covariates is updated during the sequential experiment. 

We have begun to consider how large $n_R$ grows asymptotically as a function of $\lambda$ (section \ref{subsec:theoretical}). There is a lot of theory to be done to figure out the optimal $\lambda$ to maximize estimator efficiency as functions of $n$, $p$, the variance-covariance matrix of the covariates, and how strong the signal of $f$ is to the noise. Also, perhaps a variable rule for $\lambda$ would be effective: if the sample size is large, the algorithm can afford to be conservative about the matches during the beginning of the experiment (such as waiting until $n_0$ to begin matching), but then become less conservative as time passes.

\section*{Acknowledgements}

We wish to thank Larry Brown, Andreas Buja, Dana Chandler, Dean Foster, John Horton, Stephen Kapelner, Katherine Propert,  Paul Rosenbaum, Andrea Troxel, Abraham Wyner, and Jos{\'e} Zubizaretta for helpful discussions and Hugh MacMullan for help with grid computing. Adam Kapelner also acknowledges the National Science Foundation for his graduate research fellowship that helped make this research possible.

\bibliographystyle{plainnat}\bibliography{refs}
\end{document}